\documentclass[twocolumn,aps,prl,floatfix,superscriptaddress]{revtex4-1}
\usepackage{amsmath,amssymb}
\usepackage{graphicx,float}
\usepackage{times,txfonts}
\usepackage{color}
\usepackage{multirow}
\usepackage{bbold}
\usepackage{color}
\usepackage{soul}
\usepackage{hyperref}
\usepackage{times,txfonts}
\usepackage{natbib}
\usepackage{nicefrac}
\usepackage{blindtext}
\usepackage[export]{adjustbox}
\usepackage{textcomp}

\newcommand{\qo}[1]{``#1''}

\renewcommand{\epsilon}{\varepsilon}
\renewcommand{\phi}{\varphi}

\newcommand{\degree}{$^{\circ}$}
\newcommand{\lP}{\mathbf{e}_\text{L}}
\newcommand{\rP}{\mathbf{e}_\text{R}}
\newcommand{\hP}{\mathbf{e}_\text{H}}

\begin{document}

\title{Arbitrary optical wavefront shaping via spin-to-orbit coupling}

\author{Hugo Larocque}
\affiliation{The Max Planck Centre for Extreme and Quantum Photonics, Department of Physics, University of Ottawa, 25 Templeton St., Ottawa, Ontario, K1N 6N5, Canada}

\author{J\'er\'emie Gagnon-Bischoff}
\affiliation{The Max Planck Centre for Extreme and Quantum Photonics, Department of Physics, University of Ottawa, 25 Templeton St., Ottawa, Ontario, K1N 6N5, Canada}

\author{Fr\'ed\'eric Bouchard}
\affiliation{The Max Planck Centre for Extreme and Quantum Photonics, Department of Physics, University of Ottawa, 25 Templeton St., Ottawa, Ontario, K1N 6N5, Canada}

\author{Robert Fickler}
\affiliation{The Max Planck Centre for Extreme and Quantum Photonics, Department of Physics, University of Ottawa, 25 Templeton St., Ottawa, Ontario, K1N 6N5, Canada}

\author{Jeremy Upham}
\affiliation{The Max Planck Centre for Extreme and Quantum Photonics, Department of Physics, University of Ottawa, 25 Templeton St., Ottawa, Ontario, K1N 6N5, Canada}

\author{Robert W. Boyd}
\affiliation{The Max Planck Centre for Extreme and Quantum Photonics, Department of Physics, University of Ottawa, 25 Templeton St., Ottawa, Ontario, K1N 6N5, Canada}
\affiliation{Institute of Optics, University of Rochester, Rochester, New York, 14627, USA}

\author{Ebrahim Karimi}
\affiliation{The Max Planck Centre for Extreme and Quantum Photonics, Department of Physics, University of Ottawa, 25 Templeton St., Ottawa, Ontario, K1N 6N5, Canada}
\affiliation{Department of Physics, Institute for Advanced Studies in Basic Sciences, 45137-66731 Zanjan, Iran}
\email{ekarimi@uottawa.ca}

\begin{abstract}
Converting spin angular momentum to orbital angular momentum has been shown to be a practical and efficient method for generating optical beams carrying orbital angular momentum and possessing a space-varying polarized field. Here, we present novel liquid crystal devices for tailoring the wavefront of optical beams through the Pancharatnam-Berry phase concept. We demonstrate the versatility of these devices by generating an extensive range of optical beams such as beams carrying $\pm200$ units of orbital angular momentum along with Bessel, Airy and Ince-Gauss beams. We characterize both the phase and the polarization properties of the generated beams, confirming our devices' performance.
\end{abstract}

\pacs{Valid PACS appear here}
\maketitle

\section{Introduction}
Optical beams are widely used in many areas of modern science such as microscopy, lithography, optical tweezers, and communications. The extent of these applications is essentially determined by the attributes of the employed beams such as their  polarization features, diffractive properties, and orbital angular momentum (OAM). For instance, OAM-carrying beams -- beams possessing a helical phase structure and a donut-shaped transverse intensity profile -- can be used to rotate micro-particles~\cite{he:95,padgett:11}, enhance optical resolution in stimulated emission depletion (STED) microscopy~\cite{hell:94}, and increase a communication channel's capacity~\cite{gibson:04,bozinovic:13}. Likewise, shape-invariant optical beams modulated by Airy~\cite{siviloglou:07b} and Bessel~\cite{durnin:87} functions have potential applications in optical trapping~\cite{baumgartl:08,mcgloin:05}. Finally, space-varying polarized light beams -- beams with certain polarization topologies -- have been shown to produce needle-shaped longitudinal electric and magnetic dipoles~\cite{dorn:03,wang:08} and exotic 3D polarization structures~\cite{freund:05,bauer:15}, hence their usefulness for exploring novel physical and optical phenomena. The above applications require individually specific beams characterized by well-defined intensity, phase, and polarization attributes, thus motivating the development of novel beam shaping methods and devices.

Spatial light modulators (SLMs) -- pixelated liquid crystal devices that are capable of introducing a specific phase modulation onto an incident beam -- have been widely used in different configurations to tailor certain optical phase, intensity, and polarization distributions~\cite{heckenberg:92,bolduc:13,maurer:07,galvez:12}. However, SLMs have a few drawbacks bounding them to laboratory-scale usage. To begin with, they are pixelated and limited to a certain phase modulation resolution (usually associated with 8-bit, 256 grayscale encryption). This restrains SLMs to diffractive configurations which makes them effectively bulky and lowers their efficiency. In addition, their operation is characterized by fairly low phase modulation rates ($\lesssim500$ Hz), which affect their dynamical capabilities. Similar \emph{diffraction} devices, such as deformable mirror arrays (DMA) and digital micro-mirror devices (DMD) have similar problems~\cite{hornbeck:90}. A few optical devices, such as axicons~\cite{arlt:00}, mirror-cones~\cite{mansuripur:11,kobayashi:12,radwell:16}, $q$-plates~\cite{marrucci:06}, astigmatic mode converters~\cite{allen:92}, and achromatic-OAM generators~\cite{bouchard:14} have been introduced to generate specific optical beams. However, the most efficient and compact of these devices are Pancharatnam-Berry phase optical elements (PBOEs)~\cite{marrucci:06b,bomzon:01} which have the ability to tailor optical beams through the addition of a geometric phase. In essence, they add a transverse phase profile to a beam upon flipping its circular polarization state. This phase alteration depends on the device's geometry. Only a few methods have so far been proposed to manufacture PBOEs~\cite{bomzon:01,marrucci:11,karimi:14,lin:14}. These techniques are mainly capable of generating azimuthally structured devices~\cite{marrucci:06,slussarenko:11,slussarenko:13,karimi:14}. Here, we present a novel configuration capable of manufacturing arbitrary PBOEs to demonstrate their versatility. We use them to generate OAM-carrying, Bessel, Airy, and Ince-Gauss optical beams. We also exploit our PBOEs' spin-to-orbit coupling properties through the generation of space-varying polarized beams by providing linearly polarized input beams to our devices.

\section{Theory}
Many physical entities can acquire a global phase upon being subjected to cyclic adiabatic processes. Such phases are commonly known as Pancharatnam-Berry (geometric) phases~\cite{pancharatnam:56,berry:84}. The latter of these terms is associated with the geometric transformation of the parameters defining the corresponding cyclic process. Among the many systems that can acquire a geometric phase are polarized optical beams through the adiabatic variation of their polarization states. These states can be expanded in terms of a two dimensional vector space and can thus be effectively mapped onto the surface of a unit sphere known as the Poincar\'e sphere. An arbitrary unitary transformation of an optical beam's polarization state can be performed by a sequence of (half- and quarter-) wave plates. Upon a cyclic polarization transformation along a closed path on the Poincar\'e sphere, a phase corresponding to half of the solid angle enclosed by the path will be added to the beam~\cite{bhandari:97}. 

This effect can be clearly illustrated through the example of a half-wave plate (HWP) acting upon light whose polarization is expressed in the circular polarization basis $\{\lP,\rP\}$ where $\lP$ and $\rP$ are respectively the left- and right-handed circular polarization vectors. Namely, when the half-wave plate is oriented at an angle of $\alpha$ with respect to its optical axis, the solid angle enclosed by the path of the beam's polarization vector on the Poincar\'e sphere is $\pm 4\alpha$. As a result, the beam's left-handed component becomes right-handed and acquires a phase of $2\alpha$ while the beam's right-handed component becomes left-handed and acquires a phase of $-2\alpha$. This action is described by
\begin{align}
	\label{eq:halfwaveAction}
	\begin{bmatrix} 
	\mathbf{e}_\text{L} \\
	\mathbf{e}_\text{R} \\
	\end{bmatrix} \xrightarrow{\text{HWP}_{\alpha}} 
	\begin{bmatrix}
	\mathbf{e}_\text{R} e^{+2 i \alpha} \\
	\mathbf{e}_\text{L} e^{-2 i\alpha} \\
	\end{bmatrix}.
\end{align}

Besides half-wave plates, which are defined by a uniform optical axis and add a constant phase to a beam, there are other types of devices, such as PBOEs, designed to add a space-varying geometric phase to an incident beam. Due to the identical nature of the phase added by both of these optical elements, one can readily establish an analogy between the optical properties of a half-wave plate and those of a PBOE. In essence, much like how the uniformity of the phase added by a half-wave plate arises from its uniform optical axis, the spatial dependence of the phase added by a PBOE can be seen as stemming from a non-uniform optical axis distribution. More specifically, this distribution is given by $\alpha(x_1,x_2)$, where $x_1$ and $x_2$ are coordinates defining a position on the beam's transverse profile. As in the case of a half-wave plate, the corresponding added geometric phase is given by $\pm2\alpha(x_1,x_2)$. However, in most cases, PBOEs are not perfect half-wave retarders, i.e. they are not defined by an optical retardation of $\pi$, but rather by a parameter $\delta$. Hence, it follows that a PBOE's unitary action $\hat{\text{U}}_q$ on an incident light beam in the circular basis, $\lP$ and $\rP$, for a given optical retardation $\delta$ and a general optical axis distribution $\alpha(x_1,x_2)$ is provided by~\cite{karimi:09}
\begin{align}
	\label{eq:unitaryAction}
\hat{\text{U}}_q\cdot \begin{bmatrix} 
	\mathbf{e}_\text{L} \\
	\mathbf{e}_\text{R} \\
	\end{bmatrix} = \cos{\left(\frac{\delta}{2} \right)}\begin{bmatrix}
	\mathbf{e}_\text{L} \\
	\mathbf{e}_\text{R} \\
	\end{bmatrix} +  i \sin{\left(\frac{\delta}{2}\right)} \begin{bmatrix}
	\mathbf{e}_\text{R} e^{+2 i \alpha(x_1,x_2)} \\
	\mathbf{e}_\text{L} e^{-2 i\alpha(x_1,x_2)} \\
	\end{bmatrix}.
\end{align}
Therefore, any beam that may be generated by imparting a space-varying phase to its transverse profile can be readily produced by PBOEs characterized by an appropriate optical axis distribution. Some distributions can in certain cases be more conveniently expressed in specific coordinate systems ($x_1,x_2$) such as Cartesian ($x,y$), cylindrical ($\rho,\phi$), and elliptical ($u,v$) systems. In the following, we provide several examples of existing and potential PBOEs that generate beams whose mathematical formulations are more easily expressed with a certain coordinate system. These beams include OAM-carrying and Bessel beams (cylindrical), Ince-Gauss beams (elliptical), and Airy beams (Cartesian).

PBOEs have so far been mostly designed to make devices known as $q$-plates \cite{marrucci:06b}. $q$-plates, a special class of PBOEs, possess an azimuthally symmetric optical axis given by $\alpha(\rho,\phi)=q\phi+\alpha_0$, where $q$ is either a half or a full integer, $\alpha_0$ is an arbitrary constant, and $\rho$ and $\phi$ are the radial and azimuthal cylindrical coordinates, respectively~\cite{marrucci:06}. These devices are of interest because of their ability to generate beams carrying a phase of $\ell \phi$, where $\ell$ is an integer, thus causing the beam to carry OAM~\cite{marrucci:11}. It has also been demonstrated that $q$-plates can be used to generate space-varying polarized light beams by either feeding linearly polarized light to the device or by adequately tuning its optical retardation \cite{cardano:12,cardano:13}.

Adding a linear radial phase dependence, e.g. $k_\rho \rho$, to these OAM carrying beams will simulate passing the beam through an axicon. The beam's resulting total phase, $k_\rho \rho + \ell \phi$, corresponds to that of a Bessel beam. The latter's transverse field is proportional to a $J_\ell(k_\rho \rho)\,\exp{(i \ell \phi)}$ term, where $\ell$ is an integer, $J_\ell$ is the $\ell^{\text{th}}$ order Bessel function of the first kind, and $k_\rho$ is the radial component of the beam's wavevector~\cite{durnin:87a}. A potential design for a PBOE that generates such a beam could be defined by an optical axis satisfying $2\alpha(\rho,\phi)=k_\rho \rho + \ell \phi$. Similar to the previous case, different circular polarization components will acquire opposite values of OAM. However, one circular polarization component will be attributed with a phase of $+k_\rho \rho$ while the other with a phase of $-k_\rho \rho$. Effects related to these radial phase components will be more thoroughly discussed later on in the following sections.

Bessel and OAM beams can be readily generated by imposing a cylindrically symmetric phase onto an incident Gaussian beam. Other types of beams require the impartment of phase patterns that are more conveniently expressed in other coordinate systems. Such beams include the elliptically symmetric Ince-Gauss (IG) beams. IG beams form an additional complete set of orthonormal solutions to the paraxial wave equation expressed in elliptic coordinates ($u,v$) \cite{bandres:04ince}. They include an Ince polynomial in their formulation defined by two integer indices $p$ and $m$, along with an additional parameter $\epsilon$ referred to as the beam's ellipticity. These polynomials are categorized either as even or odd Ince polynomials and correspondingly yield the so called even, $\text{IG}^e_{p,m,\epsilon}$, and odd, $\text{IG}^o_{p,m,\epsilon}$, Ince-Gauss modes respectively. The number of hyperbolic and elliptic nodal lines in the transverse intensity profile of both sets of modes is determined by their $m$ and $p$ indices. 

Beams resulting from the superposition of even and odd IG modes with a relative phase difference of $\pi/2$, i.e., $\text{IG}^{\pm}_{p,m,\epsilon} = \text{IG}^e_{p,m,\epsilon} \pm  i\,\text{IG}^o_{p,m,\epsilon}$ ($\pm$ sign denotes the direction along which the beam's phase varies), are called helical IG beams and exhibit optical vortices and an elliptic phase dependence \cite{bandres:04helical}. Helical IG modes were used to investigate several quantum features including two- and three-dimensional entanglement between their constituent photons \cite{krenn:13} and hence have potential applications in quantum information protocols \cite{plick:13}. 

PBOEs characterized by an optical axis satisfying the relation $2\alpha(u,v)=\Phi_\text{IG}(u,v)$, where $\Phi_\text{IG}$ is the phase of a helical IG beam, would be able to generate beams defined by interesting polarization properties. For instance, consider the case where $\Phi_\text{IG}$ is associated with an $\text{IG}^{-}_{p,m,\epsilon}$ beam. According to Eq.~(\ref{eq:unitaryAction}), the beam resulting from feeding an arbitrarily polarized Gaussian beam into a half-wave retarding ($\delta=\pi$) IG PBOE will consist of two components, namely $\exp{(i\Phi_\text{IG})} \rP$ and $\exp{(-i\Phi_\text{IG})}\, \lP$. As previously mentioned, the sign in the formulation of helical IG modes defines the direction along which the beam's phase elliptically varies. In our case, adding a phase of  $-\Phi_\text{IG}$ would be equivalent to generating the $\text{IG}^{+}_{p,m,\epsilon}$ helical IG mode. With these factors brought into consideration, we can equivalently write the two components of the device's output beam as $\text{IG}^{-}_{p,m,\epsilon}\rP$ and $\text{IG}^{+}_{p,m,\epsilon}\, \lP$. Therefore, by feeding linearly polarized light into the device, the relative amplitude of both components should be equal from which it follows that the resulting beam can be decomposed into two orthogonally polarized linear components, one of which being the helical IG mode's \emph{even} mode and the other being the corresponding \emph{odd} mode.

Other types of beams, such as Airy beams, require the addition of a phase that can most readily be expressed in Cartesian coordinates. Like Bessel beams, Airy beams are diffraction-free solutions to the paraxial wave equation~\cite{balazs:79}. One of their distinguishing characteristics is the fact that their main intensity maxima seem to freely accelerate along a parabolic trajectory. To produce these beams, one can exploit the fact that the Fourier transform of an Airy function $\text{Ai}$ is proportional to a term with a phase varying with the cube of the frequency~\cite{siviloglou:07b,siviloglou:07}. Thus, adding a phase with a cubic spatial dependence on an incident Gaussian beam and then taking its Fourier transform using an appropriate lens leads to a lab-scale realization of the corresponding Airy beam~\cite{siviloglou:07b} within a certain propagation range. For instance, the following space-varying phase pattern can be used to generate a 2D Airy beam
\begin{align}
	\label{eq:AiryPhase}
	\Phi_{\text{Ai}}(x,y) = \frac{1}{3}\left\{\left[x_0\left(\frac{2\pi}{\lambda f}x\right)\right]^3+\left[y_0\left(\frac{2\pi}{\lambda f}y\right)\right]^3\right\} ,
\end{align}
where $f$ is the focal length of the spherical converging lens used to take the Fourier transform of the beam, $\lambda$ is the beam's wavelength, and $x_0$ and $y_0$ are parameters that describe the beam's deflection along the $x$ and $y$ coordinates, respectively, upon propagation. As in the case of the aforementioned PBOEs, an Airy beam generated by a PBOE is defined by certain polarization features. In essence, the left-handed component of an incident beam will acquire a phase of $\Phi_{\text{Ai}}(x,y)$ while the right-handed component a phase of $-\Phi_{\text{Ai}}(x,y)$. An appealing feature of this particular device is the fact that its added phase is an odd function satisfying $-\Phi_{\text{Ai}}(x,y)=\Phi_{\text{Ai}}(-x,-y)$. This odd symmetry causes the left handed component of the converted beam to look inverted with respect to the right handed component, which leads to an \qo{acceleration} along the opposite direction.

\section{Fabrication and Characterization}
\begin{figure}[t]
\begin{center}
\includegraphics[width=0.9\columnwidth]{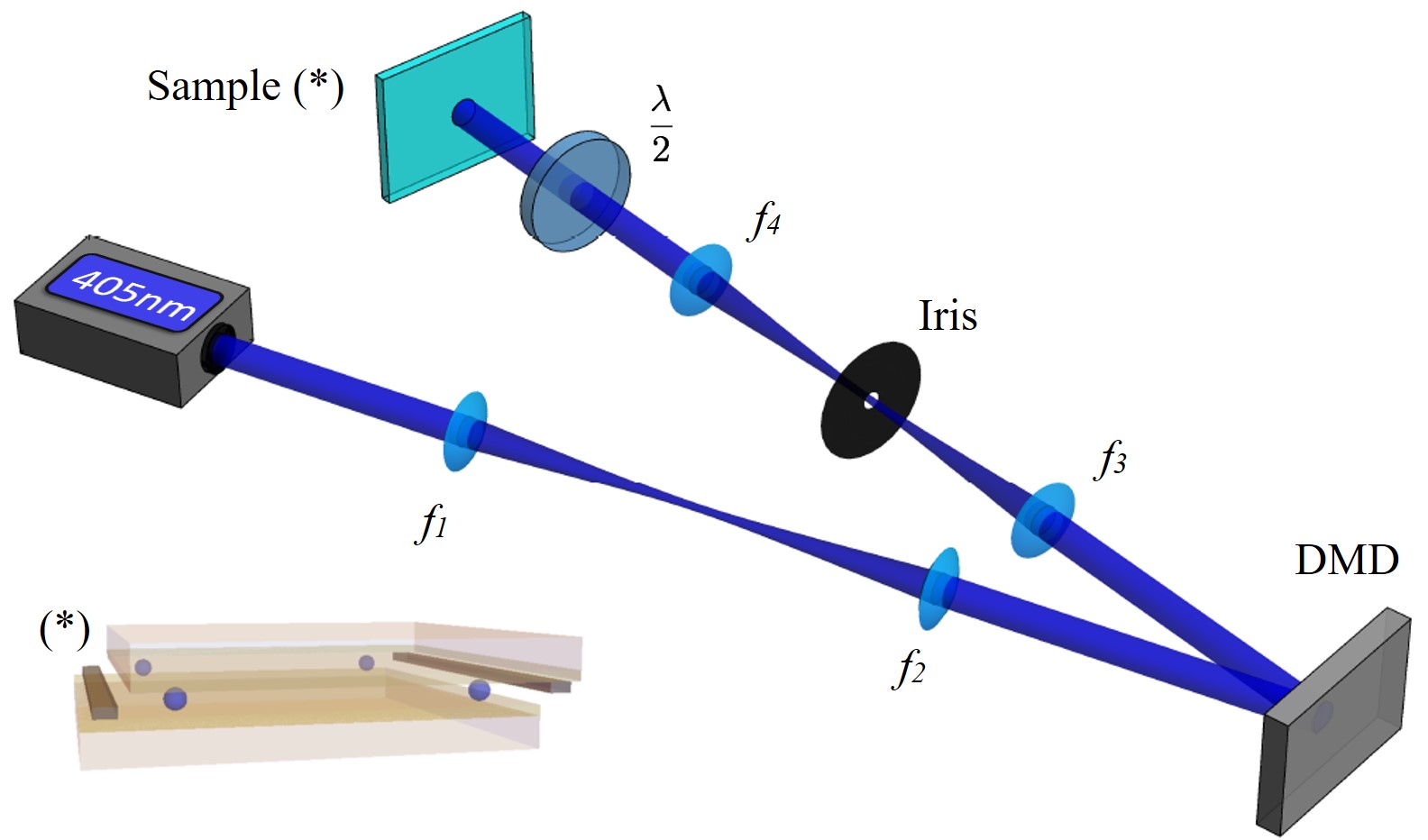}
\caption[]{Fabrication apparatus. A 405 nm linearly polarized light beam is rotated and enlarged by going through a cylindrical 4-f system ($f_1$ and $f_2$). The beam then interacts with the DMD programmed to reflect a specific light intensity pattern. The reflected beam goes through another 4-f system ($f_3$ and $f_4$) and an iris in order to select only its brightest diffraction orders as a mean to clearly image the pattern onto the sample. The polarization of the beam is then adjusted by a half-wave plate to the required orientation. After this rotation, the resulting beam aligns a specific region of the sample's photo-alignment layer. The device's structural frame is depicted in the inset in the lower-left region of the figure.}
\label{fig:apparatus}
\end{center}
\end{figure}

Cylindrically symmetric PBOEs can be fabricated by a variety of methods such as through the use of sub-wavelength gratings~\cite{bomzon:01,lin:14}, total internal reflection~\cite{bouchard:14}, plasmonic metasurfaces~\cite{karimi:14,bouchard:14a}, or an inhomogeneous and anisotropic layer of patterned nematic liquid crystals~\cite{marrucci:06b}. Here, we employ the last of these methods to produce our PBOEs designed to add an arbitrary geometric phase to optical beams.

We first describe the method to construct a PBOE with a desired optical-axis pattern imprinted into its liquid crystal layer. Such a construction can be achieved through the photoalignment of an azobenze-based dye. To perform this alignment, the azodye molecules must be exposed to specific linearly polarized light intensity patterns whose wavelength lies within a peak of the dye's absorption spectrum, thus aligning the molecules with the pattern's polarization~\cite{slussarenko:11,slussarenko:13,nersisyan:09}. When such an aligned azodye molecule is in contact with the PBOE's liquid crystals, an electrostatic interaction between the two types of molecules will align the liquid crystal molecules along the axis of the azodye. In practice, the contact between the two molecules is achieved by spin coating an azodye layer onto the inner region of two substrates enclosing the device's liquid crystals. Thus, by exposing well-defined areas of the coated substrates to specifically polarized intensity patterns, one is able to construct the PBOE's space dependent optical axis. This design can also be modified to allow one to control the PBOE's optical retardation. More specifically, when substrates coated with a layer of a transparent conductive material such as indium tin oxide (ITO) are used, electrical wires can be soldered onto these substrates. This design modification allows one to apply an electric field to the PBOE's liquid crystals, thus giving the liberty to control its optical properties including its birefringence. As a result, the element's optical retardation becomes tunable, hence allowing one to use the device to generate a wider selection of optical beams and providing the ability to use the device on beams with different wavelengths \cite{slussarenko:11}.

We produce the device's dye-coated substrates by spin-coating the azobenzene-based aligning material PAAD-22 provided by BEAM Co onto the conductive side of an ITO-coated substrate using the procedure described in~\cite{nersisyan:09}. More specifically, the sample is spin coated for 60 s at a rotational speed of 3000 rpm and then soft baked at 120\degree C for 5 minutes. This procedure typically results in the formation of a 50-nm-thick alignment layer, as we verified by observation using a profilometer. Before being exposed to light patterns, the two coated substrates are attached together with the coated region facing towards the inside of the device using an epoxy glue. We glued them with a slight lateral offset in order to ensure that wires can be soldered onto the inner conductive sides of the substrates for tuning. Spacer-grade silica microspheres with diameters below 4.8 \textmu m placed between the two plates ensure the presence of a uniformly spaced cavity in which liquid crystals can later be added. Prior to permanently gluing the plates together, the cavity's uniformity is visually optimized by minimizing the number of fringes produced by thin film interference resulting from the space between the plates. At this stage of the fabrication process, the device consists of the PBOE's structural frame and is depicted in the inset of Fig.~\ref{fig:apparatus}.

The current state-of-the-art method used to fabricate liquid-crystal-based PBOEs can only be used to fabricate devices with an optical axis that solely depends on the azimuthal coordinate. It specifically relies on rotating the sample under an angular light intensity pattern of a distinct polarization rotating at a different frequency than that of the sample~\cite{slussarenko:11,nersisyan:09}. In essence, the resulting alignment will depend on the rotation frequency ratio of these two elements.

To generate more complex optical beams, we use a photoalignment method similar to~\cite{wu:12} able to fabricate PBOEs characterized by an arbitrarily distributed optical axis. Our method relies on discretizing a PBOE's optical axis pattern into $n$ segments each of which is associated with a single orientation. Therefore, we reconstruct this discretized version of the PBOE by sequentially exposing linearly polarized light patterns to specific regions of the alignment layer defined by the corresponding optical axis orientation. This method can be implemented via a DMD photoalignment method where the DMD is used to generate the required optical patterns. As displayed in Fig.~\ref{fig:apparatus}, the rate at which the DMD produces varying light patterns is then synchronized to the rate at which a half-wave plate rotates the polarization of a linearly polarized light beam. Doing so allows the reconstruction of the PBOE's desired optical axis onto its photoalignment layer. Additional systems of lenses are added to the setup in order to refine the quality of the alignment. The first system consists of a set of two cylindrical lenses effectively rotating the beam by $90^\circ$ and enlarging its size by a factor of three in order to make it uniformly cover the DMD's display. The second set of lenses consists of a 4-f system of two spherical lenses imaging the pattern reflected by the DMD onto the sample's alignment layer while enlarging it by a factor of 2.5 thus exposing the PBOE's frame to a rectangular beam with a length of 0.9 cm and a width of 1.6 cm. Moreover, using an iris, only the brightest diffraction orders of the pattern reflected by the DMD are selected in order to increase the resolution of the image projected on the sample. After the azodye layers are aligned, liquid crystals are inserted into the structural frame's cavity which is later sealed using epoxy glue. Finally, wires are soldered on the edge of each plate using an alloy composed of 97\% indium and 3\% silver.

\begin{figure}[t]
\begin{center}
\includegraphics[width=0.9\columnwidth]{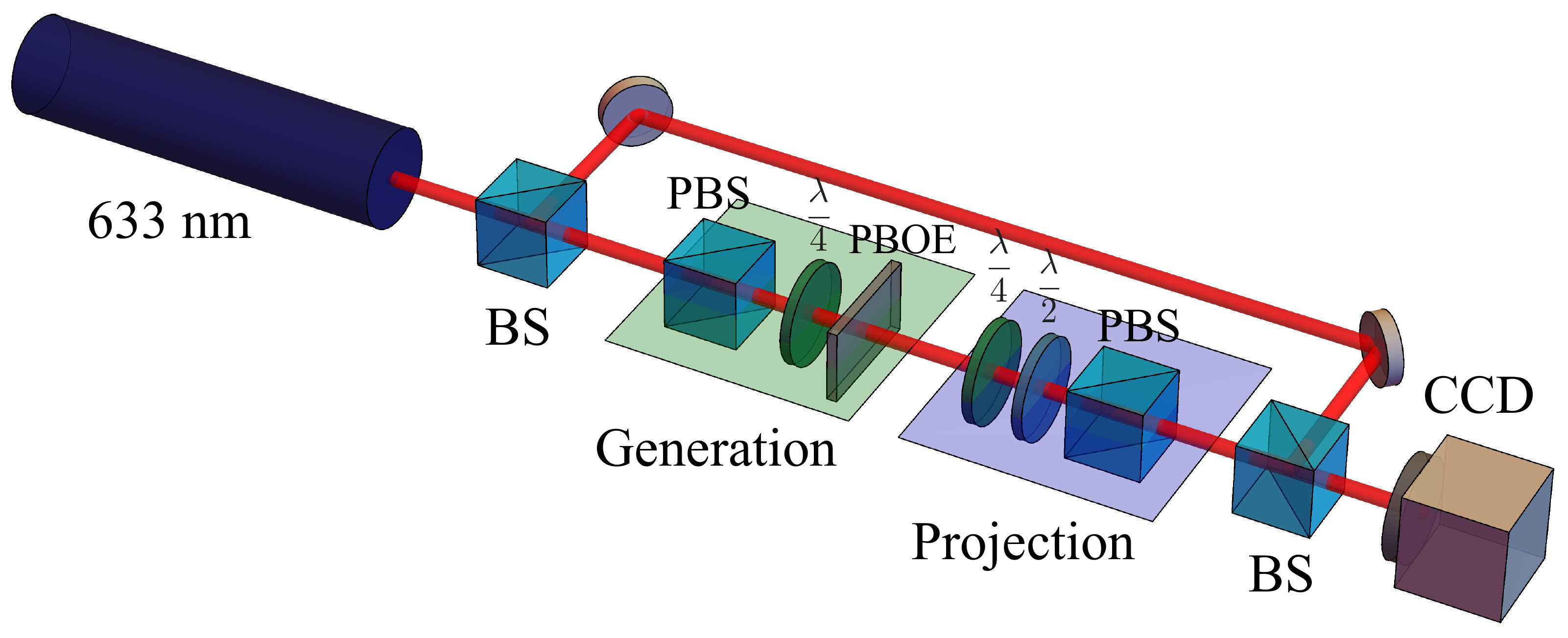}
\caption[]{Characterization apparatus that includes a 633 nm laser beam going through a Mach-Zehnder-like interferometer where the desired beam is generated and its polarization is controlled. The group of elements that generates a specific beam includes a PBS, a quarter-wave plate ($\lambda/4$), and a PBOE; the group of elements that select a single polarization includes a quarter-wave plate, a half-wave plate ($\lambda/2$), and a PBS. The beam resulting from this interferometer is then observed using a CCD camera.}
\label{fig:characSetup}
\end{center}
\end{figure}

In order to fully characterize the beams generated by the manufactured PBOEs (transverse intensity, phase, and polarization profiles), we investigate the modulation of a beam with a Gaussian profile and a wavelength of 633~nm with the setup shown in Fig.~\ref{fig:characSetup}. Prior to modulating the beam with the device, we modify its polarization by using a polarizing beam splitter (PBS) and a quarter-wave plate (QWP, $\lambda/4$). This PBS-QWP combination allows us to send either horizontally or circularly polarized light through the PBOE. Hence, we are able to generate several sorts of beams along with their associated superposition combinations. After the desired beam is generated, it goes through a QWP, HWP ($\lambda/2$), and PBS combination to investigate the various polarization components of the generated transverse beam's profile~\cite{cardano:12}. Finally, a reference Gaussian beam, which was split from the beam by a non-polarizing beam splitter before the first QWP, is made to interfere with one of the output beam's polarization components, thus making its phase profile become observable in the resulting interference pattern. The reference beam can be blocked when the setup needs to be used only to investigate the intensity profile of the beams generated by the PBOEs. The resulting intensity or interference profiles are then recorded with a CCD camera. 

The devices' transmission efficiencies were also evaluated using this apparatus. These efficiencies usually go up to 72\%. However, most losses are associated with an incident beam's reflections on the device's glass substrates and can be reduced by applying an anti-reflection coating on the optical element. Doing so would allow the PBOE's transmission efficiency to reach a value closer to its conversion efficiency, which can reach values as high as 97\%.

\section{$q$-plates}
\subsection{$q=1$- plate}

\begin{figure}[h!]
\begin{center}
\includegraphics[width=0.9\columnwidth]{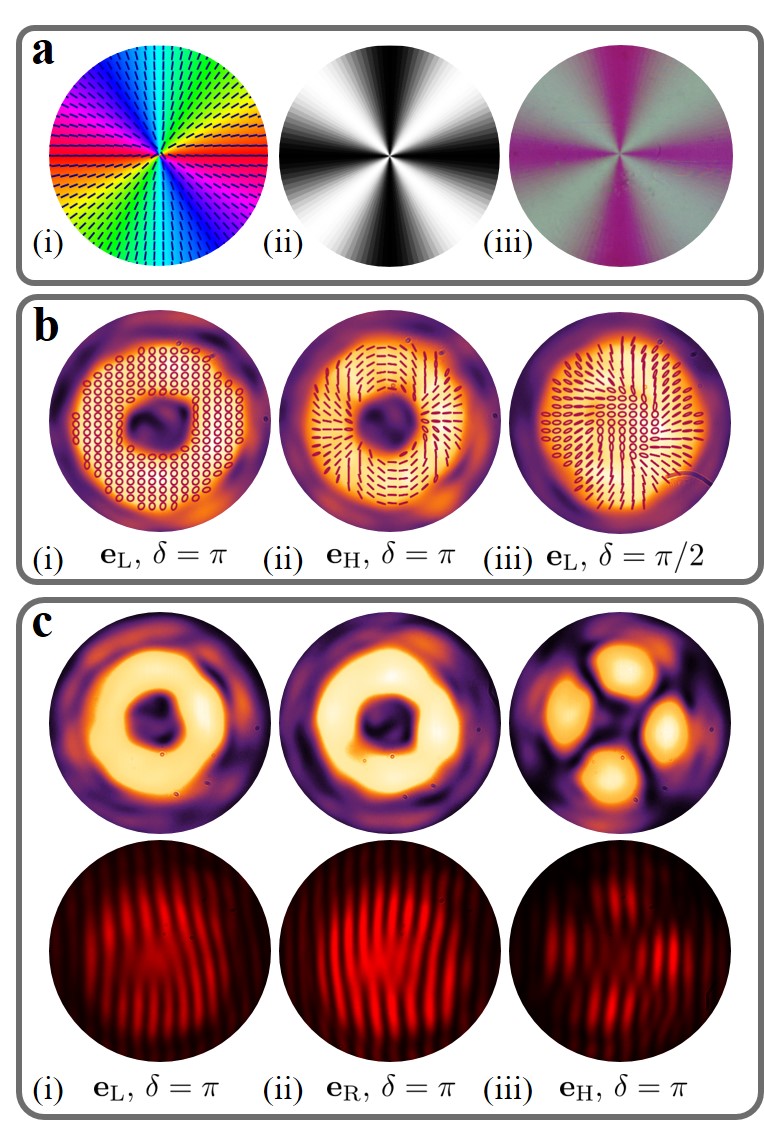}
\caption[]{\textbf{a} (i) Optical axis distribution $\alpha(\rho,\phi)$ of the $q=1$-plate with the corresponding geometric phase $2\alpha(\rho,\phi)$, shown in hue colors, added onto an incoming beam; (ii) expected transmission when the sample is placed between two polarizers; (iii) observed transmission pattern. \textbf{b} Far-field intensity and polarization profiles of the beams generated using the fabricated plate. The input polarization and tuning of the $q$-plate are indicated for each case.  (i) Uniformly right-handed circularly polarized $\ell=2$ beam; (ii) \qo{dipole} vector vortex beam; (iii) \qo{spiral} Poincar\'e beam. \textbf{c} Intensities and interference patterns with a plane wave of beams generated using (i) left-handed circularly, (ii) right-handed circularly, and (iii) horizontally polarized input light beams.}
\label{fig:q1Fig}
\end{center}
\end{figure}

To test the proposed method, we fabricate a standard $q=1$-plate, which is able to generate a vortex beam with an OAM value of $\pm2$. The device was fabricated using $n=64$ pattern method and was designed to imprint a transverse phase of $\pm2\phi$ onto an incoming Gaussian beam. The device's optical axis distribution and its corresponding added geometric phase profile are shown in Fig.~\ref{fig:q1Fig}. The device's expected transmission profile when placed between two linear polarizers is also displayed along with an image of the actual sample placed between such polarizers.

As a first test, we electrically tuned the $q=1$-plate to an optical retardation of $\delta=\pi$, which, according to Eq.~(\ref{eq:unitaryAction}), results in the entire input beam obtaining a phase profile of $\pm2\phi$. This leads to a phase singularity in the beam center imprinted onto the incident Gaussian input beam. Such beams are known as hypergeometric Gaussian beams~\cite{karimi:09}. The intensity and polarization profile of a beam resulting from illuminating the tuned $q=1$-plate with circularly polarized light is shown in Fig.~\ref{fig:q1Fig}~\textbf{b}-(i). As expected, the beam is circularly polarized and possesses a donut-like intensity profile.

As investigated previously, $q$-plates can be used to generate space-varying polarized light beams, namely vector vortex~\cite{cardano:12} and Poincar\'e beams~\cite{cardano:13}. When linearly polarized light is sent through a tuned $q$-plate, a so-called vector vortex beam is produced. In the case of a horizontally polarized input and a $q=1$-plate, one obtains a vector beam defined by a linear polarization with an orientation given by twice the azimuthal coordinate. This specific beam is commonly referred to as a \qo{dipole}-polarization beam as its polarization pattern resembles the field pattern of a dipole (see Fig.~\ref{fig:q1Fig}~\textbf{b}-(ii)).

If the $q$-plate is tuned to an optical retardation of $\delta=\pi/2$, the resulting beam consists of an equally weighted superposition of a converted and a non-converted component as seen in Eq.~(\ref{eq:unitaryAction}). Such a \emph{detuned} $q$-plate's output beam is depicted in Fig.~{\ref{fig:q1Fig}}~\textbf{b}-(iii). Here, the beam consists of an equally weighted superposition of a Gaussian beam and an $\ell=2$ vortex beam. The equal composition is clearly seen through the beam's relatively flat-top central intensity that quickly decreases as it reaches the radius where the $|\ell|=2$ vortex beam's intensity profile starts to drop. This output also corresponds to the Poincar\'e beam that can be generated using a $q=1$-plate~\cite{cardano:13}. In our case, the resulting Poincar\'e beam is defined by a polarization profile resembling a spiral, a result which is indeed experimentally observed and depicted in Fig.~\ref{fig:q1Fig}~\textbf{b}-(iii).

To observe the generated beam's transverse phase structure, the reference arm in Fig.~\ref{fig:characSetup} was used to perform interferometric measurements. A right-handed circularly polarized beam characterized by the index $\ell=2$ was generated at the output of the first arm. When it was interfered with an intentionally slightly \emph{misaligned} Gaussian reference beam, the interference pattern displayed in Fig.~{\ref{fig:q1Fig}}~\textbf{c}-(i) was obtained. The misaligned interference pattern reveals a pitchfork with three tines. 
Given that the interference between a plane wave and a beam with a phase-front defined by $\ell$ intertwined helices results in a pitchfork interference pattern consisting of $\ell+1$ tines~\cite{marrucci:06,slussarenko:11,marrucci:11}, the recorded structure confirms that the beam has an $\ell$ value of 2. When a left-handed circularly polarized beam having an index of $\ell=-2$ was generated and interfered with the Gaussian reference beam, we obtained the pattern in Fig.~{\ref{fig:q1Fig}}~\textbf{c}-(ii) which exhibits a similar pitchfork of opposite orientation as it carries OAM of opposite sign.

The QWP before the $q$-plate was then oriented to provide horizontally polarized light to the $q=1$-plate, thus again generating the vector vortex beam seen in Fig.~{\ref{fig:q1Fig}}~\textbf{b}-(ii). After projecting onto the horizontal component of the resulting beam by a second PBS, we find the expected four lobes defined by two possible distinct phases in the vicinity of either 0 or $\pi$. As a result, the interference fringes observed in the region separating the lobes are out of phase as observed in Fig.~\ref{fig:q1Fig}~\textbf{c}-(iii).

All tests examining the intensity, phase, and polarization profiles of beams generated using the manufactured $q=1$ plate reveal that the properties of these beams are in excellent agreement with their corresponding expected and well-established results.

\subsection{High OAM device: $q = 100$-plate}

We further demonstrate the method's ability to fabricate highly complex $q$-plates by manufacturing a device defined by a high topological charge, namely a $q=100$-plate able to generate optical beams characterized by $\pm200$ units of OAM. Devices able to generate extreme high-order modes are good demonstrations of the ability to explore the unbounded nature of OAM. They can also be used to produce photonic gears which can potentially be useful in optical/quantum metrology~\cite{d:13,fickler:12}. Due to the relatively high resolution needed to fabricate the device associated with the high frequency at which its optical axis varies, the $q$-plate was made using a lower reconstruction resolution consisting of a four step discretization method.  The high degree at which the $q$-plate's optical axis orientation varies results in the formation of a Moir\'e pattern in the plate's added phase. Thus, for practical reasons, the device's center is refrained from being exposed to an incident light beam. This technique is often employed in many areas of beam shaping involving the fabrication of elements defined by abrupt variations in the phase aimed to be added to a beam~\cite{fickler:12,grillo:15}

\begin{figure}[h]
\begin{center}
\includegraphics[width=0.8\columnwidth]{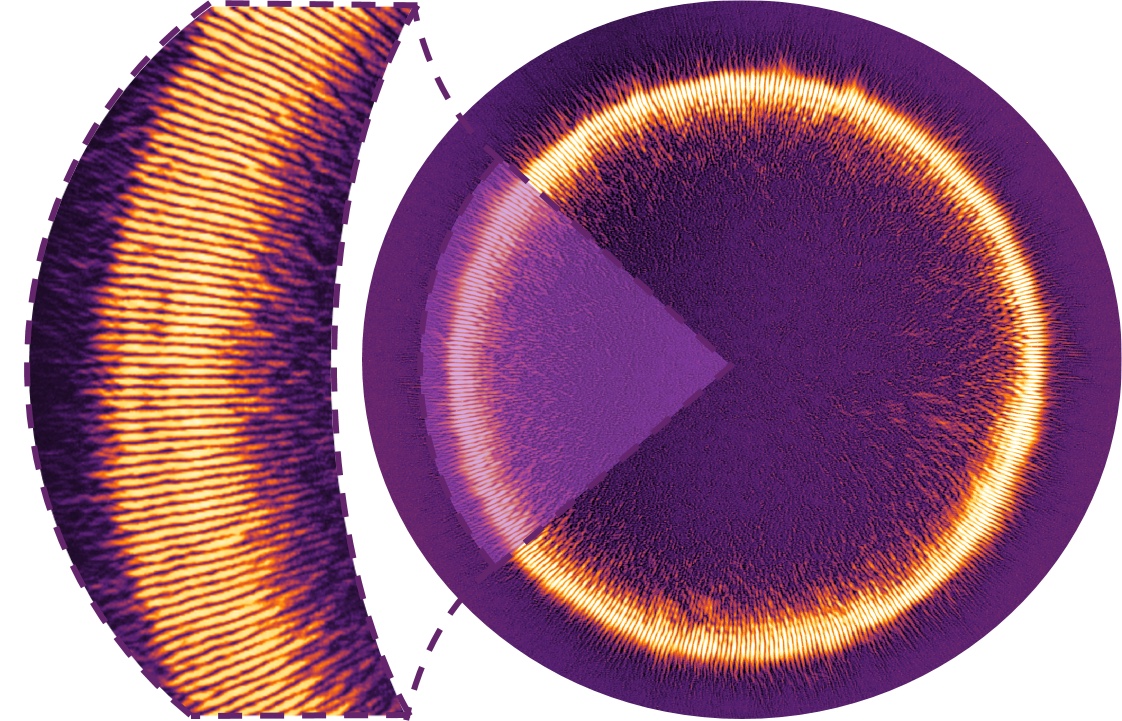}
\caption[]{A 400 petal beam generated using a $q=100$-plate fabricated using the procedure described in the text.}
\label{fig:photonicGear}
\end{center}
\end{figure}

We demonstrate the $q$-plate's performance by feeding horizontally polarized light to it and only selecting the generated beam's horizontal polarization component, which should consist of $400$ lobes. As depicted in Fig.~\ref{fig:photonicGear}, all 400 lobes of the generated beam are present, clearly defined, and uniformly shaped across the beam's profile.
\section{Generation of complex structured beams}
\subsection{Bessel Beams}

\begin{figure}[h]
\begin{center}
\includegraphics[width=0.9\columnwidth]{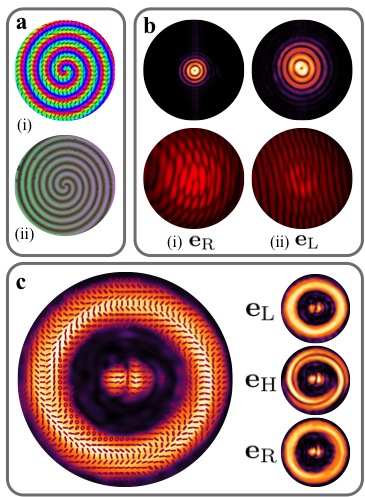}
\caption[]{ \textbf{a} (i) Optical axis distribution $\alpha(\rho,\phi)$ of a PBOE generating a Bessel beam defined by parameters of $k_\rho= 5~{(\text{mm})}^{-1}$ and $\ell = 1$ along with the corresponding geometric phase $2\alpha(\rho,\phi)$, which is shown in hue colors; (ii) its observed transmission between two polarizers. \textbf{b} Experimentally observed intensity and interference pattern of the (i) right- and (ii) left-handed circular components of the generated beam at each of their foci. \textbf{c} Reconstructed polarization profile of a generated beam using the Bessel PBOE. Measured intensities for $\lP$, $\hP$, and $\rP$ components are depicted on the right side.}
\label{fig:besselFig}
\end{center}
\end{figure}

After verifying that our method works, we continue using it to fabricate an extensive range of PBOEs in a way that has not been possible before. We first fabricate a PBOE generating a Bessel beam defined by parameters of  $k_\rho= 5~{(\text{mm})}^{-1}$ and $\ell = 1$ using a $n=32$ pattern method. 
Similarly to the $q$-plate described above, we show our device's optical axis distribution along with the corresponding geometric phase modulation as well as its observed appearance between two crossed polarizers in Fig.~\ref{fig:besselFig}~\textbf{a}. 

As Bessel beams exhibit their defining characteristics at their focus, we modified the characterization apparatus (depicted in Fig.~\ref{fig:characSetup}) to first enlarge the beam and illuminate a wider region of the device to then weakly focus it using a 1000~mm lens. Thus, we are able to better resolve the characteristics in its focal region. The right-handed and left-handed circular components of this beam have a phase with a radial dependence given by $k_\rho \rho$  and $-k_\rho \rho$, thus it will exhibit converging and diverging lensing behaviors, respectively. As a result, the Bessel beam attributed to the right-handed component will be seen to focus before the one attributed to the left-handed component. Figure \ref{fig:besselFig}~\textbf{b} shows the intensity profile of both circular polarizations at their respective foci. Since $\rP$ focuses closer to the plate, it forms a tighter Bessel beam than the one formed by $\lP$.

To examine the phase of the generated beams, we recorded interference patterns resulting from the interference between the converted circular components of the beams and a reference beam with a uniform phase front (see Fig.~\ref{fig:besselFig}~\textbf{b}). The interference patterns exhibit circular discontinuities, which are attributed to the fact that each ring appearing in a Bessel beam's intensity profile is out of phase with its adjacent rings. Moreover, these patterns are defined by a pitchfork consisting of two tines, which is a sign that the beam carries an absolute OAM value of 1. In our experimental images these features are clearly observed for both components of the generated Bessel beam. In addition, the different orientation of the pitchfork for each component demonstrates the opposite sign of their OAM quanta.

Finally, we explore the ability to use the fabricated device as a mean to generate space-varying polarized light beams. In particular, we examine the polarization of an emitted beam consisting of equally weighted circular polarization components in a region where both components perfectly overlap. The resulting beam along with its reconstructed complex polarization profile is shown in Fig.~\ref{fig:besselFig}~\textbf{c}. In addition to azimuthal variations in the polarization pattern attributed to the beam's azimuthally dependent phase, one can also see the effects of the radial dependence of the beam's phase through radial variations in the beam's transverse polarization profile. This feature can be made more visible if the beam is projected onto different polarization components. Namely, both circular components ($\lP$ and $\rP$) have a uniform appearance over the beam's cross-section, yet the linear component $\hP$ is found to form intertwined spirals.

\subsection{Ince-Gauss Beams}
\begin{figure}[h]
\begin{center}
\includegraphics[width=0.9\columnwidth]{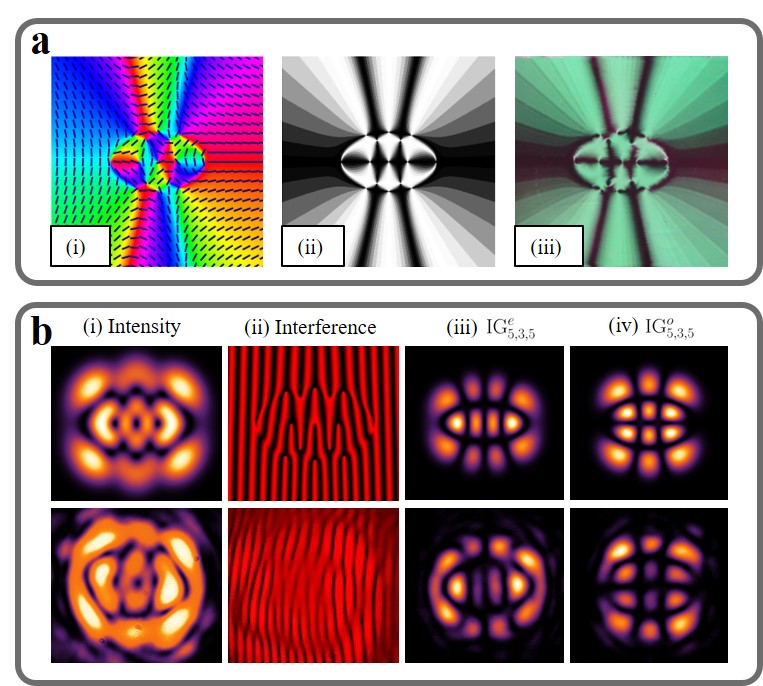}
\caption[]{\textbf{a} (i) Optical axis distribution $\alpha(u,v)$, and the corresponding geometric phase $2\alpha(u,v)$ shown in hue colors; (ii) expected device appearance between two crossed polarizers; (iii) image of the $\text{IG}^{-}_{5,3,5}$ PBOE. \textbf{b} Expected (top) and experimental (bottom): (i) intensity patterns of the completely converted component of the beam produced by the fabricated $\text{IG}^{-}_{5,3,5}$ PBOE; (ii) corresponding fringes resulting from the interference of this beam with a plane wave. Theoretical intensity profiles of the (iii) even and (iv) odd IG modes and their recovered experimental intensity profiles.}
\label{fig:IG53Plate}
\end{center}
\end{figure}

Using our custom alignment method, we were able to fabricate a device producing an $\text{IG}^{-}_{5,3,5}$ beam. Again, we used an $n=32$ pattern method in the fabrication procedure to discretize the required continuous modulation. In Fig.~\ref{fig:IG53Plate}~\textbf{a} the PBOE's desired optical axis along with the corresponding added geometric phase are shown. The device's expected appearance when placed between two crossed linear polarizers is also displayed together with an image of the actual sample.
The IG-PBOE's characterization was performed similarly to the previous ones. The expected and experimentally recorded intensities and interference patterns with a plane wave were examined and are shown in Fig.~\ref{fig:IG53Plate}~\textbf{b}. To verify the polarization properties of the generated beams, we feed linearly polarized light into the perfectly tuned device and recover the resulting beam's even IG mode and its odd IG mode. As each of these modes is attributed to a linear polarization component, we were able to recover them by accordingly modifying the orientation of the setup's half-wave plate. These recovered beams along with their corresponding theoretical transverse intensity profiles are shown in Fig.~\ref{fig:IG53Plate}~\textbf{b}.

\subsection{Airy Beams}
\begin{figure}[h]
\centering\includegraphics[width=0.9\columnwidth]{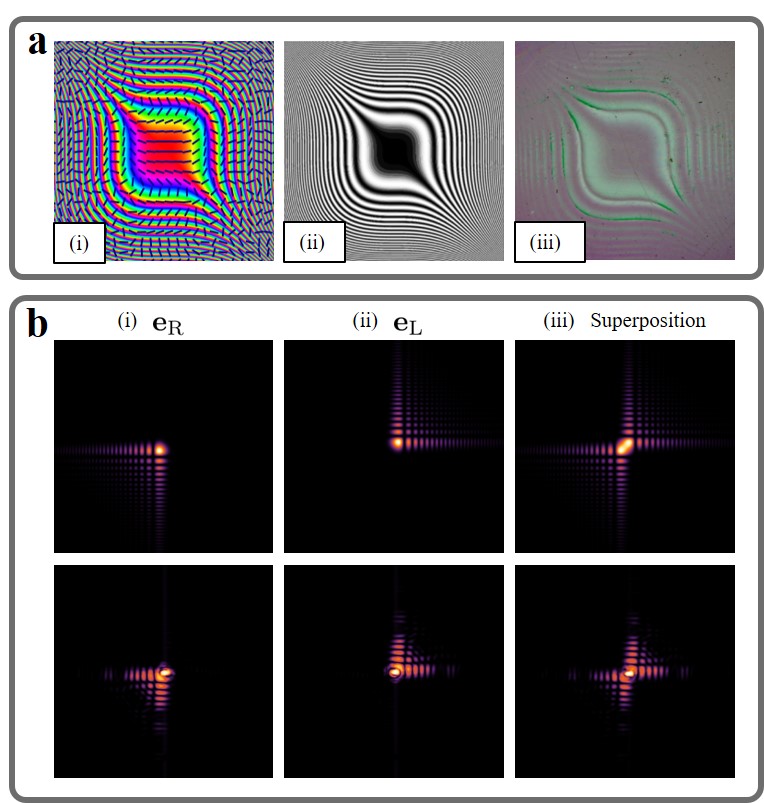}
\caption{\textbf{a} (i) Optical axis distribution $\alpha(x,y)$ and the corresponding geometric phase $2\alpha(x,y)$, which is shown in hue colors; (ii) expected appearance between two crossed polarizers; (iii) image of the device generating a 2D Airy beam. \textbf{b} Expected (top) and observed (bottom) intensity profiles of the (i) right-handed circular polarization $\rP$, (ii) left-handed circular polarization $\lP$, and (iii) a superposition of both components of the generated Airy beam.}
\label{fig:AiryPlate}
\end{figure}

As a final demonstration of the flexibility of our fabrication method, we reproduce the fabrication of PBOEs producing Airy beams as performed in \cite{wei:15}. More specifically, we manufactured a device designed to add a phase of $\Phi_{\text{Ai}}(x,y)$ defined in Eq.~(\ref{eq:AiryPhase}). The employed parameters were $f=500~mm$, $\lambda=633~nm$, $x_0=y_0=60 ~\mu m$. Similar to the earlier findings, we display in Fig.~\ref{fig:AiryPlate} the PBOE's optical axis distribution and the corresponding geometric phase along with its expected and observed appearance between two crossed polarizers.
Each polarization component generated by such a device will appear to be inverted with respect to the other. This inversion is clearly depicted in the intensity patterns of the generated beams shown in Fig.~\ref{fig:AiryPlate}~\textbf{b} along with their corresponding expected appearance, which are in excellent agreement.

\section{Conclusion}
In conclusion, we have devised and implemented a DMD-based photoalignment method to fabricate liquid crystal Pancharatnam-Berry phase optical elements defined by arbitrary optical axis spatial distributions. These PBOEs are more compact, cost-effective, and responsive to phase inversions than other beam shaping devices restricted to laboratory-scale usage. In particular, we manufactured devices capable of generating OAM-carrying, Bessel, Ince-Gauss, and Airy beams. Moreover, our generation of complex patterns with high resolution displays our ability to produce arbitrary beam patterns. These complex beams are characterized by distinct transverse intensity, phase, and polarization profiles along with the high conversion efficiency required for many applications such as quantum information and optical tweezers. For instance, cascading these PBOEs along with suitably fast Pockels cells may be used for ultra-fast (at gigahertz rate) and super-dense telecommunications. 

\section*{Funding Information}
Canada Research Chairs; Canada Excellence Research Chairs (CERC); Government of Canada; Natural Sciences and Engineering Research Council of Canada (NSERC).

\section*{Acknowledgments}
The authors thank Sergei Slussarenko and Svetlana Lukishova for useful advice and discussions.

\end{document}